\documentclass[12pt]{amsart}
\usepackage[dvips]{color}
\usepackage{amsmath}
\usepackage{amsxtra}
\usepackage{amscd}
\usepackage{amsthm}
\usepackage{amsfonts}
\usepackage{amssymb}
\usepackage{eucal}
\usepackage{epsfig}
\usepackage{graphics}
\usepackage{epsf,graphics,mathrsfs,yfonts,eufrak}

\def\'{\char126}
\def\`{\char127}

\def\({\left(}
\def\){\right)}

\newcommand{\cb}{\mathbf{c}}
\newcommand{\bb}{\mathbf{b}}

\newcommand{\gb}{\mathbf{g}}
\newcommand{\jb}{\mathbf{j}}

\newcommand{\nn}{\nonumber}

\newcommand{\slt}{\mathfrak{sl}_2}

\newcommand{\Tr}{{\rm Tr}}

\newenvironment{tenumerate}{
  \begin{enumerate}
  
  }{\end{enumerate}}
\newcommand{\bi}{\begin{tenumerate}}
\newcommand{\ei}{\end{tenumerate}}
\newcommand{\isoto}[1][]%
{{\mathop{\buildrel{\sim}\over\longrightarrow}\limits_{#1}}}

\def\[{\left[}
\def\]{\right]}
\newcommand{\la}{\lambda}

\newcommand{\al}{\alpha}

\numberwithin{equation}{section}

\def\xb{\mathbf{x}}

\def\half{\textstyle{\frac  1 2}}

\def\bi{\mathbf{i}}

\usepackage[dvipsnames]{xcolor}

\begin{document}

\begin{title}[Fermion-current basis and correlation functions]{
Fermion-current basis and correlation functions for the integrable spin 1 chain}
\end{title}
\date{\today}
\author{C.~Babenko and  F.~Smirnov}
\address{CB, FS\footnote{Membre du CNRS}: 
 Sorbonne Universite, UPMC Univ Paris 06\\ CNRS, UMR 7589, LPTHE\\F-75005, Paris, France}\email{cbabenko@lpthe.jussieu.fr,smirnov@lpthe.jussieu.fr}

\begin{abstract}

We use the fermion-current basis in the space of local operators for {the} computation of 
{the} expectation values for the integrable spin chain of spins 1. Our main tool consists in  expressing a given
local operators in the fermion-current basis. For this we use the same method as in the spin 1/2 case
which is based on {the} arbitrariness of the Matsubara data.

\end{abstract}

\maketitle

\section{Introduction}

In the papers \cite{DiFSm,MiSm} 
a method was described which allows to compute expectation values of
local operators (up to 11 sites long) for the spin 1/2 isotropic spin chain . This method is based on the 
results of the paper \cite{HGSIII} in which for the six-vertex model
(possibly inhomogeneous one) the expectation values of local operators 
in the fermionic basis are computed in terms of a function $\omega$ defined by the Matsubara data. 
Let us describe briefly the method of \cite{DiFSm,MiSm}. 

Every local operator allows a decomposition on the fermionic basis with the coefficients  depending only on 
the operator in question. For sufficiently simple Matsubara data the expectation value of the
operator can be computed in two ways: directly with the help of the algebraic Bethe ansatz and using the decomposition on the fermionic
basis and the function $\omega$. This provides {equations} for the coefficients of the
decomposition for any given Matsubara data. Repeating this procedure for sufficiently large
number of Matsubara data one obtains a system of equations for the coefficients
which allow to {find} them. 

In the present paper we apply {a} similar method to {the} much more complicated case of 
the integrable isotropic spin-1 chain described by  the Hamiltonian
\begin{align}
H=\sum _{j=-\infty}^{\infty} \(S_j^aS_{j+1}^a-(S_j^aS_{j+1}^a)^2\)\,,\label{ham}
\end{align}
where summation over $a$ is implied, $S^a$ are generators of {the} spin-1 representation of $\slt$,
{whose expression will be given below}.
The infinite chain (called Space below) is understood as {the} limit of finite chains with periodic
boundary conditions. 

The correlation functions for the model 
\eqref{ham}
 were studied in \cite{klumper}.  Later in \cite{JMSspin1} the problem was considered
 in the spirit of {the} fermionic basis construction \cite{HGSIII}. The authors of  \cite{JMSspin1} were very much influenced by \cite{klumper}.
 In the present paper we use \cite{klumper} in two ways: indirectly through \cite{JMSspin1}, and directly,
 comparing exact results on 2 and 3 sites.

Below we formulate our problem. The exposition is close to that of  the paper \cite{DiFSm} where some more 
details can be found.

The integrable models are closely related with  Quantum Groups, in particular the isotropic
model under consideration is related to the $\slt$ Yangian. We denote by $\pi_S$ the representation
obtained as {the} tensor product of the spin-1 representations along the Space.
In addition we introduce a finite, possibly inhomogeneous and {carrying} different spins,
Matsubara chain and corresponding representation $\pi_M$ of the Yangian. 
{We visualise the lattice  on an infinite cylinder with the compact direction been the Matsubara space.}
The
fundamental object is the evaluation of the universal R-matrix $\mathcal{R}$:
$$\mathbf{T}_{S,M}=\(\pi _S\otimes \pi _M\)\mathcal{R}\,.$$

The relation with the integrable spin chain is due to the commutativity
$$[H,\Tr_M(\mathbf{T}_{S,M})]=0\,,$$
which reflects the fact that $H$ is just one element of a huge commutative algebra generated by the transfer-matrices 
$\Tr_M(\mathbf{T}_{S,M})$ computed for all possible
Matsubara chains.

Denote by $\mathrm{Md}$ the data for a given Matsubara chain (length, spins, inhomogeneities).
For a  local operators {$\mathcal{O}$} localised (acting non-trivially) on a finite subchain of the Space chain,
define the expectation value
\begin{align}
\langle\mathcal{O}\rangle_\mathrm{Md}=\frac{\Tr_S\Tr_M\(\mathbf{T}_{S,M}\cdot \mathcal{O}\)}{\Tr_S\Tr_M\(\mathbf{T}_{S,M}\)}\,.\label{expectation}
\end{align}

Using the results of  the paper \cite{JMSspin1} it can be
shown that there is a basis of the local operators for the spin-1 chain  created by {the} action on the unit operators by
two  fermions and one Kac-Moody current (details will be given in the main text). 
We shall call this {the} fermion-current basis.
Denote the elements of the fermion-current  basis by
$v_\al$. For any $\mathcal{O}$ we have
$$\mathcal{O}=\sum_\al X_\al v_\al\,,$$
{where $X_\al$ are the wanted coefficients of the decomposition depending on the inhomogeneities of the Space.}
This implies
$$\langle\mathcal{O}\rangle _\mathrm{Md}=\sum_\al X_\al \langle v_\al\rangle _\mathrm{Md}\,.$$
For reasonable simple Matsubara data there are independent ways to compute
$\langle\mathcal{O}\rangle _\mathrm{Md}$ and $\langle v_\al\rangle _\mathrm{Md}$. This is how we get equations for $X_\al$.

\section{Basis}

\subsection{Homogeneous case}
We begin this section {by} making our notations more detailed.
We have the $\slt$ Yangian $\mathcal{Y}$. Denote by $\pi^{2s}_\la$ the $(2s+1)$-dimensional evaluation representation
with the evaluation parameter $\la$.
{
In order to handle $\mathbf{T}_{S,M}$ in the definitions above,  we use the following expression for the $R$ matrix of the spin 1 chain 
 $R(\lambda,\mu) = (\pi_\lambda^2\otimes \pi_\mu^2) \mathcal{R} $
which depends only on the difference of arguments $\zeta=\lambda-\mu$ :
\begin{tiny}
\begin{equation}
\nonumber
R(\zeta) =
\left(
\begin{array}{ccccccccc}
 (\zeta +1) (\zeta +2) & 0 & 0 & 0 & 0 & 0 & 0 & 0 & 0 \\
 0 & \zeta  (\zeta +1) & 0 & 2 (\zeta +1) & 0 & 0 & 0 & 0 & 0 \\
 0 & 0 & (\zeta -1) \zeta  & 0 & 4 \zeta  & 0 & 2 & 0 & 0 \\
 0 & 2 (\zeta +1) & 0 & \zeta  (\zeta +1) & 0 & 0 & 0 & 0 & 0 \\
 0 & 0 & \zeta  & 0 & \zeta +\zeta^2+2 & 0 & \zeta  & 0 & 0 \\
 0 & 0 & 0 & 0 & 0 & \zeta  (\zeta +1) & 0 & 2 (\zeta +1) & 0 \\
 0 & 0 & 2 & 0 & 4 \zeta  & 0 & (\zeta -1) \zeta  & 0 & 0 \\
 0 & 0 & 0 & 0 & 0 & 2 (\zeta +1) & 0 & \zeta  (\zeta +1) & 0 \\
 0 & 0 & 0 & 0 & 0 & 0 & 0 & 0 & (\zeta +1) (\zeta +2) \\
\end{array}
\right)
\end{equation}
\end{tiny}
}

In the homogeneous case 
$$\pi_S=\cdots \pi_0^2\otimes \pi_0^2\otimes \pi_0^2\otimes \pi_0^2\otimes \cdots\,.$$
As has been said we are supposed to begin with a finite periodic Space chain of length $2N$ and then
consider the limit $N\to\infty$. However, it is well-known that in the cylindric geometry adopted in this paper the limiting
procedure is trivial, so, we shall consider the Space chain as an infinite one. 
There is a well-known infinite family of commuting local integrals of motion which includes the Hamiltonian. 
The adjoint action of these operators is well-defined on the space of local operators. We denote by $\mathcal{V}$ the corresponding quotient space.
For the problem considered in this paper this is the space of interest.

The simplest operator $I$ acts as {a} unit  operator
in every tensor component. In \cite{JMSspin1} several operators were introduced acting on the space of local operators, let
us describe them. We start with the operators 
$\mathbf{j}^-(\la)$, $\mathbf{j}^0(\la)$,
$\mathbf{j}^+(\la)$, $\bb^*(\la)$, $\cb^*(\la)$,
for which we shall often use the universal notation
$\mathbf{x}^{\{1,2\}}=\bb^*$, $\mathbf{x}^{\{2,1\}}=\cb^*$, { $\mathbf{x}^{\{1,3\}}=\mathbf{j}^+$,
$\mathbf{x}^{\{2,2\}}=\mathbf{j}^0$, $\mathbf{x}^{\{3,1\}}=\mathbf{j}^-$. }
The indices ${\{1,2\}}$ {\it etc.} are natural in the framework of \cite{JMSspin1}.
All these operators are understood as generating functions
$$\mathbf{x}^\epsilon(\la)=\sum\limits _{p=-\infty}^{\infty}\la ^{p-1}\mathbf{x}^\epsilon_p\,.$$
It is almost correct that the space $\mathcal{V}$ is created by action of $\mathbf{x}^\epsilon_p$ with $p>0$, but some
refinements are needed. {The} first of them concerns the normal ordering. 
The operators $\mathbf{j}^-(\la)$, $\mathbf{j}^0(\la)$,
$\mathbf{j}^+(\la)$ form an {$\widehat{\slt}$} Kac-Moody algebra at level $1$. The fermions $\bb^*(\la)$, $\cb^*(\la)$ form an
{$\widehat{\slt}$} doublet. That leads to {the} natural commutation relations and, most importantly
for our goals, to the rules of the normal ordering:
\begin{align}
&:\mathbf{j}^0(\la)\mathbf{j}^0(\mu): = \mathbf{j}^0(\la)\mathbf{j}^0(\mu) - \frac{2}{(\la-\mu)^2} \,,
\quad
:\mathbf{j}^+(\la)\mathbf{j}^-(\mu): = \mathbf{j}^+(\la)\mathbf{j}^-(\mu) + \frac{\mathbf{j}^0(\mu)}{\la-\mu} +\frac 1{(\la-\mu)^2}\,,
\nn\\&:\mathbf{j}^+(\la)\mathbf{j}^0(\mu): = \mathbf{j}^+(\la)\mathbf{j}^0(\mu) + \frac{2\mathbf{j}^+(\mu)}{\la-\mu} \,,
\quad
:\mathbf{j}^0(\la)\mathbf{j}^-(\mu): = \mathbf{j}^0(\la)\mathbf{j}^-(\mu) + \frac{2\mathbf{j}^-(\mu)}{\la-\mu} \,,
\nn\\&
:\mathbf{b}^*(\la)\mathbf{j}^-(\mu):= \mathbf{b}^*(\la)\mathbf{j}^-(\mu)-\frac{\mathbf{c}^*(\mu)}{\la-\mu}\,,
\quad
:\mathbf{c}^*(\la)\mathbf{j}^+(\mu):= \mathbf{b}^*(\la)\mathbf{j}^-(\mu)+\frac{\mathbf{b}^*(\mu)}{\la-\mu}\,,
\nn\\&
:\mathbf{b}^*(\la)\mathbf{j}^0(\mu):= \mathbf{b}^*(\la)\mathbf{j}^0(\mu)+\frac{\mathbf{b}^*(\mu)}{\la-\mu}\,,
\quad
:\mathbf{c}^*(\la)\mathbf{j}^0(\mu):= \mathbf{c}^*(\la)\mathbf{j}^0(\mu)-\frac{\mathbf{c}^*(\mu)}{\la-\mu}\,.\nn
\end{align}
So, the local operators are created {by} acting on the unit operator by normal {ordered} products
\begin{align}
:\mathbf{x}^{\epsilon_1}_{p_1}\cdots \mathbf{x}_{p_l}^{\epsilon_l}:I\,,\qquad p_j>0\nn\,.
\end{align}
Introduce the ordering { $\{1,2\}\prec\{2,1\}\prec\{1,3\}\prec\{2,2\}\prec\{3,1\}$. }
For the sake of definiteness we shall require $\epsilon_1\preceq \epsilon_2\le \cdots \preceq  \epsilon_l$.
The second problem is that of completeness. Contrary to the case of {the} spin 1/2 chain \cite{completeness}
we do not have a formal proof of {the} completeness in the present situation. On the other hand
the Russian doll construction discussed below makes the completeness quite plausible. 

Let us discuss now the most complicated issue. An  important question is that of how the
operators located exactly on the interval $[1,n]$ look like in our fermion-current  basis. In the spin 1/2 case we had
only fermionic operators $\bb^*_p,\cb^*_p$. For the
operators 
\begin{align}  \bb^*_{p_1}\cdots\bb^*_{p_k}\cb^*_{q_1}\cdots \cb^*_{q_l} I\,,\label{bc}\end{align}
to be located on $[1,n]$ one imposes first of all two necessary  conditions:
\begin{align}
&1)\qquad k+l\le n\,,\label{1cond}\\ \  \nn\\
&2)\qquad  p_j\le n,\  q_j\le n\ \ \ \forall j\,. \label{2cond}
\end{align}
Then there are more subtle necessary conditions explained in details in 
\cite{DiFSm,MiSm}. Taking into account all the necessary conditions we come to the subspace
of the fermionic space, whose elements may be located on $[1,n]$, of rather reasonable size.
Notice also that in \cite{DiFSm,MiSm} as well as in the present paper we are interested in operators 
 invariant under the action of global $\slt$. This requires $k=l$ in \eqref{bc}. 

For the spin 1 case, let us write the element of the fermion-current  basis in complete notations
\begin{align}
 :\bb^*_{p_1}\cdots\bb^*_{p_{k_1}}\ \cb^*_{q_1}\cdots \cb^*_{q_{k_2}}
 \ \jb^+_{r_1}\cdots\jb^+_{r_{k_3}}  \ \jb^0_{s_1}\cdots\jb^0_{s_{k_4}} \ \jb^-_{t_1}\cdots\jb^-_{t_{k_5}} :I\,.\label{bcj}
\end{align}
There is one necessary condition which remains unchanged:
\begin{align}
k_1+k_2+k_3+k_4+k_5  \le n\,.\label{kkkkk}
\end{align}
{The} requirement of $\slt$-invariance of {the} operators is equivalent to
$$k_1-k_2+2k_3-2k_5=0\,.$$
For fermions the condition \eqref{2cond} and additional conditions from \cite{DiFSm,MiSm} (null-vectors) still hold.
However, we were not able to formulate reasonable conditions for the currents. That is why 
in what follows, we are forced
to take much more complicated and less efficient ways {to calculate the correlations functions of the fermion-current basis},
than in \cite{DiFSm,MiSm}.

%%%%%%%%%%%%%%%%%%%%%%%%%%%%%%%%%%%%%%%%%%%%%%%%%%%%%%%%%%%%%%%%%%%%%%%%%%%%%%%%%%%%%%%%%%%%%%%%%

\subsection{Introducing Matsubara}
The Matsubara chain is inhomogeneous
$$\pi_M=\pi_{\tau_1}^{2s_1}\otimes \pi_{\tau_2}^{2s_2}\otimes \cdots \otimes\pi_{\tau_L}^{2s_L} \,.$$
{Let us} introduce the transfer-matrix
$$\mathbf{T}_M(\la)=\(\Tr\otimes \mathrm{id}\)(\pi^{(2)}_\la\otimes \pi _M)(\mathcal{R})\,.$$
This is a commutative family, {for generic Matsubara data} there is a unique eigenvector
with the maximal in absolute value eigenvalue of $\mathbf{T}_M(0)$. We shall denote this eigenvector by $|\Psi\rangle$.
{The} corresponding eigenvalue of the transfer-matrix will be denoted by $\mathbf{T}(\la)$.

Clearly for  any local operator located on the interval $[1,n]$  we have
\begin{align}
\lim_{N\to \infty}\frac{\Tr_S\Tr_M\(\mathbf{T}_{S,M}\cdot \mathcal{O}\)}{\Tr_S\Tr_M\(\mathbf{T}_{S,M}\)}=\frac
{\langle \Psi|\Tr _{[1,n]}\(\mathbf{T}_{[1,n],M}\mathcal{O}\)|\Psi\rangle}{\mathbf{T}(0)^n\langle \Psi|\Psi\rangle}\,,
\label{topsi}
\end{align}
{where $\mathbf{T}_{[1,n],M}$ is the restriction of $\mathbf{T}_{S,M}$ for the Space  taken to be the finite interval $[1,n]$, 
its explicit expression is given below for the inhomogeneous case.}
Our way of computing the right hand side does not depend on the fact that the
eigenvalue is maximal being applicable to any eigenvector of the transfer-matrix.

\subsection{Inhomogeneous case,  Russian doll}
The Russian doll construction is present indirectly already in the paper \cite{HGSII}, however, in 
\cite{JMSspin1} it becomes really indispensable. The construction requires some definitions which we  are going to give.

We shall need an inhomogeneous space chain:
$$\pi_S=\cdots \pi_0^2\otimes \pi_0^2\otimes \pi_{\la_1}^2\otimes\cdots \otimes \pi_{\la_n}^2\otimes\pi_0^2\otimes\pi_0^2 \otimes \cdots\,.$$
The inhomogeneity is located on a finite subchain $[1,n]$. 
Consider the space {of} all the operators located on this interval.
Consider the expectation value \eqref{expectation} for the inhomogeneous case assuming that the local
operator $\mathcal{O}$ is located on the interval $[1,n]$. Denote  {the} corresponding spaces, isomorphic to $\mathbb{C}^3$ , by $V_1,\cdots ,V_n $.

In order to describe a suitable for our goals
basis in $V_1\otimes\cdots\otimes V_n$
we introduce nine operators $\gb^\epsilon(\la_k)$ ($\epsilon=\{i,j\}$, $i,j=1,2,3$) and act by these operators on $I$ consequently:
$$\gb^{\epsilon_n}(\la_n)\gb^{\epsilon_{n-1}}(\la_{n-1})\cdots \gb^{\epsilon_1}(\la_1)I\,.$$
For generic $\la_1,\cdots,\la_n$ this gives a basis of the space of  operators localised on the interval $[1,n]$. 
{We have the equality} $\gb^{\{1,1\}}(\la)=\mathrm{id}$. The expectation values considered in the present paper are such that in the weak sense
{(holding when considered in correlation functions)}
\begin{align} 
\gb^{\{3,3\}}(\la)%\raisebox{.2cm}{${\textstyle \mathrm{w}}$} 
{\ }^{{}^\mathrm{w}}
\hskip -.4cm
= \gb^{\{1,1\}}(\la)\,,\quad \gb^{\{2,3\}}(\la){\ }^{{}^\mathrm{w}}
\hskip -.4cm
= \gb^{\{1,2\}}(\la)\,,\quad \gb^{\{3,2\}}(\la){\ }^{{}^\mathrm{w}}
\hskip -.4cm
= \gb^{\{2,1\}}(\la)\,.\label{ggg9to5}
\end{align}
So, effectively we are left with the same set of indices counting the operators $\gb$ as we had before for $\xb$.

As usual the monodromy matrix $(\pi^2_{\la_j}\otimes \pi_M)(\mathcal{R})$ with the first tensor
component identified with $V_j$ will be devoted by $\mathbf{T}_{j,M}(\la_j)$.
The formula \eqref{topsi} remains valid for $\mathcal{O}$ being located
on the interval $[1,n]$, and, certainly,
$$\mathbf{T}_{[1,n]}=\mathbf{T}_{1,M}(\la_1)\cdots \mathbf{T}_{n,M}(\la_n)\,.$$
 These operators {$\gb^\epsilon$} are in one-to-one correspondence 
with $\xb$'s. Wanting to pass to the homogeneous case one has {to} apply the 
normal ordering, the rules are the same as above. The Russian doll construction
is based on the identity
\begin{align}
&\lim_{N\to \infty}\frac{\Tr_S\Tr_M\(\mathbf{T}_{S,M} :\xb^{\epsilon_n}(\la_n)\cdots \xb^{\epsilon_1}(\la_1):I\)}{\Tr_S\Tr_M\(\mathbf{T}_{S,M}\)}\label{hominhom}\\&=\frac
{\langle \Psi|\Tr _{[1,n]}\(\mathbf{T}_{1,M}(\la_1)\cdots \mathbf{T}_{n,M}(\la_n)\ :\gb^{\epsilon_n}(\la_n)\cdots \gb^{\epsilon_1}(\la_1):I\)|\Psi\rangle}{\prod_{j=1}^n\mathbf{T}(\la_j)\langle \Psi|\Psi\rangle}\,.\nn
\end{align}
This formula establishes {an} identity between the expectation values of a family of local operators of different
lengths for the homogeneous case with the expectation values for the operators of length $n$ in the inhomogeneous case. 
For our goals, rather complicated reasonings concerning this formula which are given in \cite{JMSspin1} can
be avoided just by saying that the explicit computation of the right hand side
(which will be given soon for any Matsubara data), defines the operators $\xb$ in the left hand side.

Still there is another way to apply this formula. Suppose one computes the right hand side and then sets all
$\la_j$ to zero. In that case the right hand side gives the expectation value of a local operator located on $[1,n]$ for
the homogeneous chain, this allows to identify the local operators of length $n$ in the left hand side.
We shall explain how to apply this idea in practice later.

\subsection{Fusion}
Consider the tensor product of $2n$ two-dimensional spaces $v_j$. 
Introduce the projector $\mathcal{P}_j\ :\ v_{2j-1}\otimes v_{2j}\to V_j$ onto the symmetric component. 
Consider the product $\mathcal{P}=\mathcal{P}_1\otimes\cdots\otimes \mathcal{P}_n$\,.
Denote by $T_{j,M}(\la)$ the monodromy matrix whose first tensor component acts in $v_j$. 
We have the fusion
\begin{align}
T_{1,M}(\la_1-1/2)T_{2,M}(\la_1+1/2)&\cdots T_{2n-1,M}(\la_n-1/2) T_{2n,M}(\la_n+1/2)\mathcal{P}\nn\\&=\mathcal{P}\ \mathbf{T}_{1,M}(\la_1)\cdots \mathbf{T}_{n,M}(\la_n)\,.\nn
\end{align}

We began to consider the tensor product of $2n$ spaces $v_j$ isomorphic to $\mathbb{C}^2$.
In the framework of the present paper the interest of this consideration is due to
the fact that 
$\pi_{\la_1}^2\otimes\cdots\otimes  \pi_{\la_n}^2$ is a submodule of
$\pi_{\la_1-1/2}^1\otimes\pi_{\la_1+1/2}^1\otimes\cdots\otimes \pi_{\la_n-1/2}^1\otimes\pi_{\la_n+1/2}^1$.
In what follows it will be useful to consider {a} more general
module $\pi_{\mu_1}^1\otimes\cdots\otimes\pi_{\mu_{2n}}^1$ with generic $\mu_1,\cdots \mu_{2n}$  specialising 
{to} $\mu_j=\la_{\[\frac {j+1} 2\]}+\frac{(-1)^j } 2$
when needed. 
We have operators  $g^\sigma(\mu_j)$ ($\sigma=\{1,2\},\ \{2,1\}$) acting on the latter space. The Matsubara expectation values are
computed via a particular case of the main fermionic basis formula \cite{HGSIII}:
\begin{align}
&\frac
{\langle \Psi|\Tr _{[1,2n]}\({T}_{1,M}(\mu_1)\cdots {T}_{2n,M}(\mu_{2n})\ g^{\sigma_{2n}}(\mu_{2n})\cdots g^{\sigma_1}(\mu_{1})I\)|\Psi\rangle}{\prod_{j=1}^n {T}(\mu_j)\langle \Psi|\Psi\rangle}\label{main1/2}\\
&=(-1)^{\mathrm{sgn}(\pi)}
\det\left|\omega(\mu_i,\mu_j)\right|_{i: \sigma_i=\{2,1\} ,\ j: \sigma_j=\{1,2\}}\,,
\nn
\end{align}
where $\pi$ is the permutation putting all $i$ such that $\sigma_i=\{2,1\}$ to the left. The functions $\omega(\la,\mu)$ depends on the Matsubara data as on parameters. We shall not repeat the definition which can be found in  \cite{DiFSm,MiSm}.

Using the formula above one computes the right hand side of \eqref{hominhom} using the following formulae
\begin{align}
&\gb^{\{1,2\}}(\la)=g^{\{1,2\}}(\la+1/2)+g^{\{1,2\}}(\la-1/2)\,,\label{gfat}\\
&\gb^{\{2,1\}}(\la)=g^{\{2,1\}}(\la+1/2)+g^{\{2,1\}}(\la-1/2)\,,\nn\\
&\gb^{\{1,3\}}(\la)=g^{\{1,2\}}(\la+1/2)g^{\{1,2\}}(\la-1/2)\,,\nn\\
&\gb^{\{3,1\}}(\la)=g^{\{2,1\}}(\la+1/2)g^{\{2,1\}}(\la-1/2)\,,\nn\\
&\gb^{\{2,2\}}(\la)=g^{\{2,1\}}(\la+1/2)g^{\{1,2\}}(\la-1/2)+g^{\{1,2\}}(\la+1/2)g^{\{2,1\}}(\la-1/2)\,.\nn
\end{align}
It is important to notice that $\gb^{\epsilon_n}(\la_n)\cdots \gb^{\epsilon_1}(\la_1)I$ in which $\gb$ are defined by \eqref{gfat} satisfies the identity
$$\gb^{\epsilon_n}(\la_n)\cdots \gb^{\epsilon_1}(\la_1)I=\mathcal{P}\gb^{\epsilon_n}(\la_n)\cdots \gb^{\epsilon_1}(\la_1)I\,,$$
which provides {the} self-consistence of {the} fusion. 

This procedure expresses  the right hand side of \eqref{hominhom} in terms of determinants of 
matrices with the matrix elements being expressed in terms of the function
$\omega(\la,\mu)$ and the normalisation
$$\mathcal{N}(\la)=\frac{\mathbf{T}({\lambda})}{T(\la+\frac 1 2)T(\la-\frac 1 2)}\,,\nn
$$
as follows
\begin{align}
&\frac
{\langle \Psi|\Tr _{[1,n]}\(\mathbf{T}_{1,M}(\la_1)\cdots \mathbf{T}_{n,M}(\la_n)\ \gb^{\epsilon_n}(\la_n)\cdots \gb^{\epsilon_1}(\la_1)I\)|\Psi\rangle}{\prod_{j=1}^n\mathbf{T}(\la_j)\langle \Psi|\Psi\rangle}
=\prod_{j=1}^n\frac 1 {\mathcal{N}(\la_j)}\label{from1to1/2}\\ \
\nn\\
%&\mathcal{F}_{\sigma_1,\cdots,\sigma_{2n}}^{\epsilon_1,\cdots,\epsilon _n}
%\frac
%{\langle \Psi|\Tr _{[1,2n]}\({T}_{1,M}(\la_1-\frac 1 2)\cdots {T}_{2n,M}(\la_{n}+\frac 1 2)\ g^{\sigma_{2n}}(\la_n-\frac 1 2)\cdots g^{\sigma_1}(\la_1-\frac 1 2)I\)|\Psi\rangle}{\prod_{j=1}^n {T}(\la_j+\frac 1 2){T}(\la_j-\frac 1 2)\langle \Psi|\Psi\rangle}\nn\\
&\times\mathcal{F}_{\sigma_1,\cdots,\sigma_{2n}}^{\epsilon_1,\cdots,\epsilon _n}\frac
{\langle \Psi|\Tr _{[1,2n]}\({T}_{1,M}(\mu_1)\cdots {T}_{2n,M}(\mu_{2n})\ g^{\sigma_{2n}}(\mu_{2n})\cdots g^{\sigma_1}(\mu_{1})I\)|\Psi\rangle}{\prod_{j=1}^n {T}(\mu_j)\langle \Psi|\Psi\rangle}\,,\nn
\end{align}
where
\begin{align}
&\{\mu_1,\mu_2,\cdots,\mu_{2n-1},\mu_{2n}\}=\{\la_1-\half,\la_1+\half,\cdots,\la_n-\half,\la_n+\half\}\,,\nn\\
\end{align}
$\mathcal{F}_{\sigma_1,\cdots,\sigma_{2n}}^{\epsilon_1,\cdots,\epsilon _n}$ is a tensor easily read from \eqref{gfat}.

\section{Computational procedure and results}
\subsection{General procedure}
In the homogeneous case consider an operator localised on the interval $[1,n]$. 
As usual we  simplify the notations in \eqref{bcj} introducing multi-indices:
$$:\bb^*_P\ \cb^*_Q\ \mathbf{j}^+_R\ \mathbf{j}^0_S\ \mathbf{j}^-_T\ I:\,.$$
Consider an operator $\mathcal{O}$ localised on the interval $[1,n]$. Our goal is 
to find the decomposition
\begin{align}
\mathcal{O}\equiv\sum\limits_{P,Q,R,S,T}\ X_{P,Q,R,S,T}\ :\bb^*_P\ \cb^*_Q\ \mathbf{j}^+_R\ \mathbf{j}^0_S\ \mathbf{j}^-_T\ I:\,,
\label{dechom}
\end{align}
where $\equiv$ means equality in the quotient by the action of the local integrals of motion space. 
We would like  to proceed as in \cite{DiFSm,MiSm}, namely,  to use sufficiently simple Matsubara data in order to
obtain equations for the coefficients $X$ {by} computing 
% in two different ways the expectation values
{independently the expectation values of operators on the right hand side and on the left hand side}. 
However, in the present case there are some complications. {The} first is the normal ordering. {The} second is the multiplier
containing $\mathcal{N}$ in \eqref{from1to1/2}, it looks quite innocent, but actually it is not. Also, as has been discussed, 
we did not find an efficient way (similar to \cite{DiFSm,MiSm}) to restrict the number of terms in the right hand side.
With all that in mind we decided to take {a} simpler way based on the inhomogeneous chain. 

In the inhomogeneous case the analogue of \eqref{dechom} looks {like}
\begin{align}
\mathcal{O}\equiv \sum\limits_{\epsilon_1,\cdots\epsilon_n}\mathcal{X}_{\epsilon_1,\cdots\epsilon_n}(\la_1,\cdots,\la _n)
:\gb^{\epsilon_n}(\la_n)\cdots \gb^{\epsilon_1}(\la_1):I\,,\label{decinhom}
\end{align}
having in mind \eqref{ggg9to5} we reduce the indices to
$\epsilon_p=\{1,1\},\{1,2\},\{2,1\},\{1,3\},\{3,1\},\{2,2\}$ remembering that $\gb^{\{1,1\}}(\la_j)=\mathrm{id}$,
and $\equiv$ stands for equality of the expectation values for 
all Matsubara data in the geometry accepted in the present paper, in other words for the case when the left and the right
Matsubara states are equal (we denote them by $|\Psi\rangle$). This is the inhomogeneous version of the quotient by {the} action of
the local integrals. 

The computation of the expectation value of \eqref{decinhom} follows closely that explained in \cite{DiFSm,MiSm}. In the left hand side we have
\begin{align}
\frac
{\langle \Psi|\Tr _{[1,n]}\(\mathbf{T}_{1,M}(\la_1)\cdots \mathbf{T}_{n,M}(\la_n)\ \mathcal{O}\)|\Psi\rangle}{\prod_{j=1}^n\mathbf{T}(\la_j)\langle \Psi|\Psi\rangle}\label{ccc}
\end{align}
The choice of Matsubara data is explained in \cite{DiFSm}.
The numerator of this expression is a linear combination of terms of the kind 
$$\langle \Psi|\mathbf{T}_{i_1,j_1}(\la_1)\cdots \mathbf{T}_{i_n,j_n}(\la_n)|\Psi\rangle\,.$$
{
where $\mathbf{T}_{i_k,j_k}(\la_k)\in \mathrm{End}(M)$ stands for the coefficient at position $ i_k,j_k $
of $\mathbf{T}_{k,M}(\la_k) = \left( \mathbf{T}_{i_k,j_k}(\la_k) \right)_{1\leq i_k,j_k \leq3 } $.
}

Using fusion, the computations are reduced to the ones explained in details in \cite{MiSm}.
The norm $\langle \Psi|\Psi\rangle$ is computed by Gaudin formula, the eigenvalue
$$\mathbf{T}(\la)={T}(\la-\half){T}(\la+\half)-\Delta(\la)\,,$$
$\Delta(\la)$ being the quantum determinant. 

%The right hand side of \eqref{decinhom} is computed applying consequently the rules of the normal ordering,
%{that is} the formulae \eqref{gfat}, then we express the result in terms of the functions $\omega(\la,\mu)$ and $\mathcal{N}(\la)$,
%the computation of such functions is explained in \cite{DiFSm,MiSm}.

The right hand side of \eqref{decinhom} is computed applying consequently the rules of the normal ordering,
{that is} the formulae \eqref{gfat}, then we express the result in terms of the functions $\omega(\la,\mu)$ and $\mathcal{N}(\la)$.

{Notice that $\omega$ appear only in expressions of the form $\omega(\la\pm\frac{1}{2},\mu\pm\frac{1}{2})$ and as well
as $\mathcal{N}$
is computed from formulae given in \cite{DiFSm,MiSm}. However  $\omega$  need to be made  compatible with the definition of
the normal order. To this end, we introduce an auxiliary function $\varphi$ :
\begin{equation}
\varphi(z)=\frac{1}{4}\Big(- \frac{3}{z+1}-\frac{1}{z-1}+\frac{3}{z}+\frac{1}{z+2} \Big) 
\label{phi}
\end{equation}
and consider  the two  redefinitions : 
\begin{equation}
\nn
 \widetilde\omega(\la+\half,\mu-\frac{1}{2}) = {\omega}(\la+\frac{1}{2},\mu-\frac{1}{2}) + \varphi(\la-\mu)\,,
 \quad
 \widetilde\omega(\la-\frac{1}{2},\mu+\frac{1}{2}) = {\omega}(\la-\frac{1}{2},\mu+\frac{1}{2}) + \varphi(\la-\mu-1)\,,
\end{equation}
where ${\omega}$ is taken as such from \cite{DiFSm,MiSm}. }

%\begin{equation}
%\nn
% \omega(\la+\frac{1}{2},\mu-\frac{1}{2}) = \widetilde{\omega}(\la+\frac{1}{2},\mu-\frac{1}{2}) + \varphi(\la-\mu)
% \quad\quad
% \omega(\la-\frac{1}{2},\mu+\frac{1}{2}) = \widetilde{\omega}(\la-\frac{1}{2},\mu+\frac{1}{2}) + \varphi(\la-\mu-1)
%\end{equation}

%\begin{equation}
%\nn \omega(\la+\frac{1}{2},\mu+\frac{1}{2}) = \widetilde{\omega}(\la+\frac{1}{2},\mu+\frac{1}{2})
%\quad\quad
%\omega(\la-\frac{1}{2},\mu+\frac{1}{2}) = \widetilde{\omega}(\la-\frac{1}{2},\mu+\frac{1}{2})
%\end{equation}

% \begin{align}
% &\nn
% \omega(\la+\frac{1}{2},\mu-\frac{1}{2}) = \widetilde{\omega}(\la+\frac{1}{2},\mu-\frac{1}{2}) + \varphi(\la-\mu)
% \quad\quad
% \omega(\la-\frac{1}{2},\mu+\frac{1}{2}) = \widetilde{\omega}(\la-\frac{1}{2},\mu+\frac{1}{2}) + \varphi(\la-\mu-1)
%\\
%&\nn \omega(\la+\frac{1}{2},\mu+\frac{1}{2}) = \widetilde{\omega}(\la+\frac{1}{2},\mu+\frac{1}{2})
%\quad\quad
%\omega(\la-\frac{1}{2},\mu+\frac{1}{2}) = \widetilde{\omega}(\la-\frac{1}{2},\mu+\frac{1}{2})
%\end{align}

Below we give some examples  
of the expressions of the simplest elements of the fermion-current basis in terms of $\omega$, and of 
how the normal ordering works in practice :
\begin{align*}
\nn
& \langle \mathbf{b}^*(\la)\mathbf{c}^*(\mu) \rangle\\&=
\mathcal{N}(\la)\mathcal{N}(\mu)
\bigl( \widetilde\omega(\la+\half,\mu+\half )+ \widetilde\omega(\la+\frac{1}{2},\mu-\frac{1}{2})+ \widetilde\omega(\la-\frac{1}{2},\mu+\frac{1}{2})+ \widetilde\omega(\la-\frac{1}{2},\mu-\frac{1}{2})\bigr) \,,\\
&\nn \langle \mathbf{j}^+(z)\mathbf{j}^-(w) \rangle = - \mathcal{N}(\la)\mathcal{N}(\mu)
 {\begin{vmatrix}
    \widetilde\omega(\la+\frac{1}{2},\mu+\frac{1}{2})  &  \widetilde\omega(\la+\frac{1}{2},\mu-\frac{1}{2})    \\
    \widetilde\omega(\la-\frac{1}{2},\mu+\frac{1}{2}) &  \widetilde\omega(\la-\frac{1}{2},\mu-\frac{1}{2})   \\
  \end{vmatrix} } + \frac{1}{(\la-\mu)^2}\,.
\end{align*}

{This type of formulae are} easy to compute for given small Matsubara data and numerical $\la_j$. Doing that we find experimentally how many
different Matsubara data we need to get the expansion \eqref{decinhom}. Denote {by} $L$ the length of the Mastubara chain and for
$B$ the number of Bethe roots. 
For example, for the most complicated case considered
in the present paper, $n=5$,  the following stock of Matsubara data is sufficient: 22  with $L=1,B=0$, 149 with $L=2,B=0$,
25 with $L=3, B=0$, 8  with $L=2,B=1$, 35  with $L=3,B=1$, 1  with $L=4,B=2$.

Up to $n=3 $ the computation is simple. The structure of the coefficients is as follows
\begin{align}
\mathcal{X}(\la_1,\cdots,\la _n)=\prod\limits_{i<j}\ \frac 1 {(\la_{i}-\la_{j})^{d_{i,j}}}\ \frac{P(\la_1,\cdots,\la _n)}{R(\la_1,\cdots,\la _n)}\,,\label{aaa}
\end{align}
where $d_{i,j},P,R$ depend on $\epsilon_1,\cdots,\epsilon_n$, $R(0,\cdots,0)\ne 0$. 
The degrees $d_{i,j}$ are easy to find: we take all $\la$'s sufficiently distant except $\la_i$ and $\la_j$ for which we consider
two
separations, say, {$10^{-12}$, $1+10^{-13}$}. Obviously,   this allows to define  $d_{i,j}$.

Using this {as an} Ansatz in {the} general case is difficult mostly because of the denominator $R(\la_1,\cdots,\la _n)$. 
On the other hand we are not really interested in all the details of this
denominator having in mind further application to the homogeneous case. Let us explain that. 

Consider the right hand side of \eqref{decinhom}.
The normally ordered expression $:\gb^{\epsilon_n}(\la_n)\cdots \gb^{\epsilon_1}(\la_1):I$ is regular 
at the point $\la_1=0,\cdots ,\la_n=0$. The left hand side of \eqref{decinhom} does not depend on $\la$'s.
So, setting $\la_j=\epsilon\la'_j$ and sending $\epsilon$ to $0$ one concludes that
in the function 
$$F(\la_1,\cdots,\la _n)=\frac{P(\la_1,\cdots,\la _n)}{R(\la_1,\cdots,\la _n)}$$ among the terms with $\epsilon^{D}$, 
only those with $D=\sum d_{i,j}$ may
contribute. The terms with $D>\sum d_{i,j}$ vanish in the limit. The singular terms with 
$D<\sum d_{i,j}$ must vanish, this gives rise to null-operators whose expectation values vanish regardless {of} the choice of
{the} Matsubara data. 

Experiments show that $F(\la_1,\cdots,\la _n)$ is invariant under simultaneous shift of arguments.
So we need the expansion 
\begin{align}
F(\la_1,\cdots,\la _n)=\sum\limits_{{m_2,\cdots m_n}\atop{\sum m_j\le\sum d_{i,j}}}\prod_{j=2}^n(\la_j-\la_1)^{m_j}F_{m_2,\cdots m_n}\,.\nn
\end{align}
Practical computations are easier in this form: we do not need to know the denominator {$R$}. 
The computation of the Taylor series are performed taking sufficiently small $\la$'s and determining 
the Taylor coefficients {$F$ step by step}. 
The coefficients of the Taylor series grow rapidly with the length of the interval $n$, hence the inconvenience
of the present procedure: for $n=5$ we are forced to take $\la$'s of the order of $10^{-30}$. This makes computations rather slow.

Having the coefficients {$F$}, we arrive after a simple computation 
at {the} final formula \eqref{dechom}.
\subsection{Examples}

The simplest $\slt$-invariant operator of length $n$ is $\sum_{a=1}^3 S^a_1S^a_n$.

{
It is defined by :
\begin{equation}
\nonumber
\sum_{a=1}^3 S^a_1S^a_n
=\frac{1}{2} h\otimes I_{n-2} \otimes h +e\otimes I_{n-2} \otimes f + f\otimes I_{n-2} \otimes e
\end{equation}
where we have the usual $\mathfrak{sl}_2$ spin 1 operators : 
\begin{equation}
\nonumber
h= \left( {\begin{array}{ccc}
   2 & 0  & 0   \\
   0 & 0 & 0   \\
   0  & 0  & -2 \\
  \end{array} } \right)
\quad
e= \left( {\begin{array}{ccc}
   0 & 2  & 0   \\
   0  &0  & 1   \\
    0 & 0  & 0 \\
  \end{array} } \right)
\quad
f= \left( {\begin{array}{ccc}
  0  & 0 & 0   \\
    1 & 0 & 0   \\
    0 & 2  & 0 \\
  \end{array} } \right)
\end{equation}
}

For $n=2,3$ we compute
\begin{align}
\sum_{a=1}^3 S^a_1S^a_2&=- \frac{34}{3}  - 4 \bb_1^*\cb_1^*-\frac{8}{3}\jb^+_1\jb^-_1\,,\label{exps}\\
\sum_{a=1}^3 S^a_1S^a_3&= -478 +\frac{384}{5}\mathbf{b}_1^*\mathbf{c}^*_1 + \frac{176}{3}(\mathbf{b}_2^*\mathbf{c}^*_2-\mathbf{b}_3^*\mathbf{c}^*_1 )
 -\frac{13216}{15}\mathbf{j}_1^+\mathbf{j}^-_1\nn\\
&
 +\frac{1024}{15}(\mathbf{j}_2^+\mathbf{j}^-_4-\mathbf{j}_5^+\mathbf{j}^-_1-\mathbf{j}_3^+\mathbf{j}^-_3-\mathbf{j}_3^+\mathbf{j}_2^0\mathbf{j}_1^- )
  \nn\\
&+224(\mathbf{j}_3^+\mathbf{j}^-_1
-\mathbf{j}_2^+\mathbf{j}^-_2 )
+240 \mathbf{b}_1^*\mathbf{b}_2^*\mathbf{j}_1^-+\frac{832}{15}(\mathbf{b}_1^*\mathbf{b}_3^*\mathbf{j}_2^- -\mathbf{b}_2^*\mathbf{b}_3^*\mathbf{j}_1^-
-\mathbf{b}_1^*\mathbf{b}_2^*\mathbf{j}_3^-) \,.\nn
\end{align}

{The first results are derived from the inhomogeneous formula ($n=2,3$) which are presented in the Appendix.}
In the case of {the} infinite volume and zero temperature the function $\omega(\la,\mu)$ simplifies a lot.
First, in this case it depends {only on the } difference of {the} arguments: $\omega(\la,\mu)=\omega(\la-\mu)$.
Second, we have the 
functional equation
\begin{align}
\omega(\la+1)+\omega(\la)=\frac{\pi}{2\sin(\pi \la)}-\varphi(\la)\,,\label{eqom}
\end{align}
where { $\varphi$ is defined in \eqref{phi}. }
% $$\varphi(z)=\frac{1}{4}\Big(- \frac{3}{z+1}-\frac{1}{z-1}+\frac{3}{z}+\frac{1}{z+2} \Big) 
% \,.$$
The equation \eqref{eqom} is easy to solve, but actually the explicit solution is never needed
in our computations:  the final results are expressed only through {the shifted sum of two $\omega$'s in } the left hand side of  \eqref{eqom}.
This explains {why} the final results are given by sums of even powers of $\pi$ with rational coefficients.
For two and three sites we have
\begin{align}
&\langle\sum_{a=1}^3 S^a_1S^a_2\rangle=\frac{8 \pi ^2}{9}-\frac{34}{3} = -2.560351643\,,\nn\\
&\langle\sum_{a=1}^3 S^a_1S^a_3\rangle=
-478+\frac{13216 \pi ^2}{45}-\frac{224 \pi ^4}{5}+\frac{4096 \pi ^6}{2025} = 1.283223553
\,,\nn
\end{align}
in full agreement with \cite{klumper}.

We found expressions similar to \eqref{exps} for $n=4,5$ which are unfortunately
too long to be presented here. They are available upon a request. But the results for the infinite volume and zero
temperature are of reasonable size:
\begin{align}
\langle\sum_{a=1}^3 S^a_1S^a_4\rangle=
& \frac{74317166}{75}-\frac{54372392 \pi ^2}{27}+\frac{14677235264 \pi ^4}{10125}-\frac{6743857664 \pi^6}{14175}\nn\\
&+\frac{238274860288 \pi ^8}{3189375}-\frac{1509154816 \pi ^{10}}{273375}+\frac{17291214848 \pi ^{12}}{111628125} =-1.083843468\,,\nn\\
\langle\sum_{a=1}^3 S^a_1S^a_5\rangle&=
 \frac{30764875058782}{175}-\frac{5889239056193536 \pi ^2}{6615}+\frac{129766077160539584 \pi^4}{70875} \nn\\ 
&-\frac{1795332485778909184 \pi ^6}{893025}+\frac{609942688710268901888 \pi ^8}{468838125}\nn\\ 
&-\frac{6922910606153603072 \pi^{10}}{13395375} +\frac{2684747793382087192576 \pi ^{12}}{21097715625}\nn\\ 
&-\frac{339956010411039064064 \pi^{14}}{17722081125} 
+\frac{7217056126203854848 \pi ^{16}}{4219543125}\nn\\ 
&-\frac{2439025898062610432 \pi   ^{18}}{29536801875}+\frac{572648486718144512 \pi ^{20}}{344596021875}= 0.8330261734\,.\nn
\end{align}
>From the expressions above one conjectures that $\langle\sum_{a=1}^3 S^a_1S^a_n\rangle$ is a polynomial in $\pi^2$ of degree
$n(n-1)/2$ with rational coefficients. 

Having  {developed}   the fermion-current basis it is easy to compute the correlators
$\langle\sum_{a=1}^3 S^a_1S^a_n\rangle$ ($n=2,3,4,5$)
for finite temperature (like in  \cite{MiSm}), or for the generalised Gibbs ensemble. 

Another interesting application consists in the computation of the density matrix {$D(n)$}
for the interval of length $n$ in the infinite antiferromagnetic chain and of the entanglement entropy. 
Our methods of computation are far from perfection, so, we are doing much worse than in
the paper \cite{MiSm}, namely, only up to $n=4$. This is not enough to compare the entanglement entropy 
{$s(n)= - \mathrm{Tr}(D(n)\log D(n))$ } with
the CFT prediction \cite{cft}
$$s(n)\simeq\frac{c} 3\log n +a=\frac 1 2 \log n +a\,,$$
where $a$ is a non-universal constant. 
We remind {that} the scaling limit {of the model} is  described by {a} CFT with $c=3/2$. 
Still some resemblance with the scaling behaviour is already observed in the table which present {the}
results of our computations.
\begin{center}
\begin{tabular}{|l|c|r|}
\hline
$n$ & $s(n)_{\ }$ & $s(n)-\frac{1}{2}\log n$   \\
    \hline
  2 & 1.5005420731509647 &  1.153968482870992  \\
   \hline

  3 & 1.7187172552051159 &  1.169411110871061  \\
   \hline

  4 & 1.8681251161018912 &  1.174977935541946  \\
  \hline
\end{tabular}
\end{center}
 \section{Conclusion}
 
 We have shown that the fermion-current basis works for small subchains of {an} infinite
 spin 1 integrable chain. In particular, the completeness holds at least up to intervals
 of length 5. We produced exact results for {lengths}  $n=4,5$ which were
 not available previously. However, we are far from the length 11 achieved in \cite{DiFSm}.
 There are two reasons for that. First,  there is an objective reason: the model is far more complicated and the
 fermion-current basis contains much more elements than the fermionic basis
 for the spin 1/2 case. Second, there is a  subjective reason: our method of computation is not perfect, we 
 did not find how to work with the homogeneous case directly, so, we are forced to mix it with the
 inhomogeneous one, in {a} rather involved way which requires a lot of computer memory.

\section{Appendix}
Here we give formulae for the inhomogeneous case in a weak sense having in mind \eqref{ggg9to5}.
The inhomogeneities are $\la_1,\cdots,\la_n$. The coefficients do not depend on a simultaneous shift of
inhomogeneities for that reason we shall use
$$\mu_j=\la_{j+1}-\la_1\,.$$

For $n=2$ we have
\begin{tiny}
\begin{align}
& \sum_{a=1}^3 S^a_1S^a_2 \; {\ }^{{}^\mathrm{w}}
\hskip -.4cm
= \;\;
\frac{2 (17 - 6 \mu_1^2 + \mu_1^4)}{3 ( \mu_1^2-1)} +(\mu_1^2-4) \mathbf{g}^{1,2}(\la_1) \mathbf{g}^{2,1}(\la_2)  
- \frac{2}{3} (\mu_1^2-4)(\mu_1^2-1)\mathbf{g}^{1,3}(\la_1) \mathbf{g}^{3,1}(\la_2)  \nn
\end{align}
\end{tiny}

For $n=3$ we have

\begin{tiny}
\begin{align}
\nn
& \sum_{a=1}^3 S^a_1S^a_3   {\ }^{{}^\mathrm{w}} \hskip -.4cm
=
-\frac 2 {45 (\mu_1^2-1) ((\mu_1 - \mu_2)^2-1) ( \mu_2^2-1) }  
\times (-10755 + 4406 \mu_1^2 - 943 \mu_1^4 + 20 \mu_1^6 - 4406 \mu_1 \mu_2 + 1886 \mu_1^3 \mu_2 
        \nn \\&
        -  60 \mu_1^5 \mu_2 + 4241 \mu_2^2 - 2499 \mu_1^2 \mu_2^2 + 342 \mu_1^4 \mu_2^2 + 
        4 \mu_1^6 \mu_2^2 + 1556 \mu_1 \mu_2^3 - 584 \mu_1^3 \mu_2^3 - 12 \mu_1^5 \mu_2^3 - 793 \mu_2^4 
        + 
        372 \mu_1^2 \mu_2^4 + \mu_1^4 \mu_2^4  
        \nn\\&
         - 90 \mu_1 \mu_2^5 + 18 \mu_1^3 \mu_2^5 + 35 \mu_2^6 -  11 \mu_1^2 \mu_2^6)
        \nn \nn\\&
-\frac{2}{15 (\mu_1^2-1) ( 
       ( \mu_1 - \mu_2)^2-1) (\mu_1 - \mu_2)  (\mu_2^2-1 ) \mu_2 } \times
         (1360 - 416 \mu_1^2 - 182 \mu_1^4 + 
        30 \mu_1^6 + 618 \mu_1 \mu_2 + 533 \mu_1^3 \mu_2 - 155 \mu_1^5 \mu_2 
        \nn\\&
         - 618 \mu_2^2   - 259 \mu_1^2 \mu_2^2
                + 300 \mu_1^4 \mu_2^2 - 5 \mu_1^6 \mu_2^2 - 548 \mu_1 \mu_2^3 - 
        284 \mu_1^3 \mu_2^3 + 67 \mu_1^5 \mu_2^3 + 274 \mu_2^4 + 127 \mu_1^2 \mu_2^4 - 
        178 \mu_1^4 \mu_2^4 
        + 5 \mu_1^6 \mu_2^4 
       \nn\\&
             + 18 \mu_1 \mu_2^5 + 
       179 \mu_1^3 \mu_2^5 - 20 \mu_1^5 \mu_2^5 - 6 \mu_2^6 - 69 \mu_1^2 \mu_2^6 + 30 \mu_1^4 \mu_2^6 + 8 \mu_1 \mu_2^7 - 
        20 \mu_1^3 \mu_2^7 - 2 \mu_2^8 + 5 \mu_1^2 \mu_2^8))
        \gb^{1,2}(\la_1)\gb^{2,1}(\la_2)
        \nn\nn\\&
                +\frac 1 {15 (\mu_1^2-1) \mu_1  ((\mu_1 - \mu_2)^2-1) (\mu_1 - \mu_2) ( \mu_2^2-1) } \times(2720 - 
    1176 \mu_1^2 + 428 \mu_1^4 + 48 \mu_1^6 - 4 \mu_1^8 + 1176 \mu_1 \mu_2 - 856 \mu_1^3 \mu_2 
    \nn\\&
    - 144 \mu_1^5 \mu_2  + 16 \mu_1^7 \mu_2 - 832 \mu_2^2 - 773 \mu_1^2 \mu_2^2 + 584 \mu_1^4 \mu_2^2 - 
    213 \mu_1^6 \mu_2^2 + 10 \mu_1^8 \mu_2^2 + 1201 \mu_1 \mu_2^3 - 928 \mu_1^3 \mu_2^3 +
    583 \mu_1^5 \mu_2^3  
    \nn\\&
     - 40 \mu_1^7 \mu_2^3 - 364 \mu_2^4+ 840 \mu_1^2 \mu_2^4 - 611 \mu_1^4 \mu_2^4 +
    75 \mu_1^6 \mu_2^4 - 400 \mu_1 \mu_2^5 + 269 \mu_1^3 \mu_2^5
    - 85 \mu_1^5 \mu_2^5 + 60 \mu_2^6  
    -55 \mu_1^2 \mu_2^6  
    \nn\\&
    +55 \mu_1^4 \mu_2^6 + 15 \mu_1 \mu_2^7 - 
    15 \mu_1^3 \mu_2^7)
    \gb^{1,2}(\la_2)\gb^{2,1}(\la_3)\nn
   \nn\\
        & -\frac{2 }{15 (\mu_1^2-1) \mu_1  ( (\mu_1 - \mu_2)^2-1)  (\mu_2^2-1) \mu_2 }\times(1360 - 416 \mu_1^2 - 182 \mu_1^4 + 30 \mu_1^6 + 
      214 \mu_1 \mu_2 + 195 \mu_1^3 \mu_2 - 25 \mu_1^5 \mu_2 
      \nn\\& 
      - 416 \mu_2^2 + 248 \mu_1^2 \mu_2^2 
      - 25 \mu_1^4 \mu_2^2 - 5 \mu_1^6 \mu_2^2 + 195 \mu_1 \mu_2^3 + 34 \mu_1^3 \mu_2^3 - 37 \mu_1^5 \mu_2^3 - 
      182 \mu_2^4 - 25 \mu_1^2 \mu_2^4 + 82 \mu_1^4 \mu_2^4
       + 5 \mu_1^6 \mu_2^4 
      \nn\\&
      - 25 \mu_1 \mu_2^5 -   37 \mu_1^3 \mu_2^5   - 10 \mu_1^5 \mu_2^5 + 30 \mu_2^6  - 5 \mu_1^2 \mu_2^6 +   5 \mu_1^4 \mu_2^6))
      \gb^{2,1}(\la_2)\gb^{1,2}(\la_3)\nn
      \nn\\&
           -\frac {8(\mu_1^2-4) (\mu_1^2-1)} {45 (-1 +
         (\mu_1 - \mu_2)^2-1) (\mu_1 - \mu_2)^2  (\mu_2^2-1) \mu_2^2 }(-96 + 32 \mu_1^2 - 
        2 \mu_1^4 + 67 \mu_1 \mu_2 - 18 \mu_1^3 \mu_2 - 67 \mu_2^2 - 7 \mu_1^2 \mu_2^2 + 5 \mu_1^4 \mu_2^2 
        \nn\\&
        +  50 \mu_1 \mu_2^3 - 25 \mu_2^4- 25 \mu_1^2 \mu_2^4 + 30 \mu_1 \mu_2^5 - 10 \mu_2^6)
        \gb^{1,3}(\la_1)\gb^{3,1}(\la_2)
        \nn\nn\\&
        -\frac{ 2 (\mu_2^2-4)}{45 (\mu_1^2-1) \mu_1  ( (\mu_1 - \mu_2)^2-1) (\mu_1 - 
        \mu_2)^2  \mu_2 }
        \times
        (-384 + 992 \mu_1^2 - 872 \mu_1^4 + 
        306 \mu_1^6 - 44 \mu_1^8 + 2 \mu_1^10 + 268 \mu_1 \mu_2 + 987 \mu_1^3 \mu_2 
        \nn\\&
        - 602 \mu_1^5 \mu_2 +  147 \mu_1^7 \mu_2 - 8 \mu_1^9 \mu_2 - 748 \mu_2^2 + 138 \mu_1^2 \mu_2^2 + 188 \mu_1^4 \mu_2^2 - 
        178 \mu_1^6 \mu_2^2 + 12 \mu_1^8 \mu_2^2 - 1097 \mu_1 \mu_2^3
        + 274 \mu_1^3 \mu_2^3 + 
        52 \mu_1^5 \mu_2^3 
        \nn\\& 
        - 21 \mu_1^7 \mu_2^3 + 708 \mu_2^4  - 194 \mu_1^2 \mu_2^4 + 
        92 \mu_1^4 \mu_2^4 + 54 \mu_1^6 \mu_2^4 + 140 \mu_1 \mu_2^5
        - 74 \mu_1^3 \mu_2^5 - 
        78 \mu_1^5 \mu_2^5 - 104 \mu_2^6 - 20 \mu_1^2 \mu_2^6 
        \nn\\&
         + 52 \mu_1^4 \mu_2^6 + 25 \mu_1 \mu_2^7 - 
        13 \mu_1^3 \mu_2^7)  \gb^{1,3}(\la_1)\gb^{3,1}(\la_3)
        \nn\nn\\&
      -\frac{4 ( (\mu_1 - \mu_2)^2-4) }{45 (\mu_1^2-1) \mu_1 (\mu_1 - \mu_2) (\mu_2^2-1) \mu_2^2 }\times(-192 + 
        256 \mu_1^2 - 68 \mu_1^4 + 4 \mu_1^6 + 614 \mu_1 \mu_2 - 402 \mu_1^3 \mu_2 + 52 \mu_1^5 \mu_2 - 
        374 \mu_2^2 + 540 \mu_1^2 \mu_2^2
        \nn\\&
        - 142 \mu_1^4 \mu_2^2 - 860 \mu_1 \mu_2^3 + 
        561 \mu_1^3 \mu_2^3 - 55 \mu_1^5 \mu_2^3 + 354 \mu_2^4 - 497 \mu_1^2 \mu_2^4 + 
        111 \mu_1^4 \mu_2^4
        + 2 \mu_1^6 \mu_2^4 + 227 \mu_1 \mu_2^5 - 128 \mu_1^3 \mu_2^5 - 
        3 \mu_1^5 \mu_2^5 
        \nn\\&
        - 52 \mu_2^6 + 55 \mu_1^2 \mu_2^6 + 3 \mu_1^4 \mu_2^6 - 5 \mu_1 \mu_2^7 - 
        \mu_1^3 \mu_2^7)  \times  \gb^{3,1}(\la_2)\gb^{1,3}(\la_3)
        \nn\nn\\&
        + \frac{2 (\mu_1^2-4) ( \mu_2^2-4) (-26 - 7 \mu_1^2 + 
    12 \mu_1 \mu_2 + 5 \mu_1^3 \mu_2 - 7 \mu_2^2 - 10 \mu_1^2 \mu_2^2 + 5 \mu_1 \mu_2^3) }{
 15 \mu_1 ( (\mu_1 - \mu_2)^2-1) (\mu_1 - \mu_2)  \mu_2)} 
 \gb^{1,3}(\la_1)\gb^{2,1}(\la_2)\gb^{2,1}(\la_3)
 \nn\nn\\& +\frac{
 2 (\mu_1^2-4) ( (\mu_1 - \mu_2)^2-4)(-26 - 2 \mu_1^2 + 2 \mu_1 \mu_2 - 
    7 \mu_2^2 + 5 \mu_1^2 \mu_2^2 - 5 \mu_1 \mu_2^3} {
 15 \mu_1 (\mu_1 - \mu_2) (\mu_2^2-1) \mu_2}
 \gb^{2,1}(\la_1)\gb^{1,3}(\la_2)\gb^{2,1}(\la_3)\nn
\nn\\&
-\frac{
  2 (( \mu_1 - \mu_2)^2-4)   ( \mu_2^2-4) (26 + 7 \mu_1^2 - 2 \mu_1 \mu_2 + 
     5 \mu_1^3 \mu_2 + 2 \mu_2^2 - 5 \mu_1^2 \mu_2^2)}{15 (\mu_1^2-1) \mu_1  (\mu_1 - \mu_2) \mu_2}
      \gb^{2,1}(\la_1)\gb^{2,1}(\la_2)\gb^{1,3}(\la_3)\nn
     \nn\\& -\frac{
  4  (\mu_1^2-4) ( (\mu_1 - \mu_2)^2-4) ( \mu_2^2-4) (-12 + 
     \mu_1^2 - \mu_1 \mu_2 + \mu_2^2))}{45 \mu_1 (\mu_1 - \mu_2) \mu_2}\
     \gb^{1,3}(\la_1)\gb^{3,1}(\la_2)\gb^{2,2}(\la_3)\nn
\end{align}

\end{tiny}

%%%%

%%%%%

\end{document}